\pdfoutput=1
\documentclass[manuscript]{acmart}
\usepackage[utf8]{inputenc}
\def\BibTeX{{\rm B\kern-.05em{\sc i\kern-.025em b}\kern-.08emT\kern-.1667em\lower.7ex\hbox{E}\kern-.125emX}}

\setcopyright{acmcopyright}
\acmJournal{TSAS}
\acmYear{2019} \acmVolume{1} \acmNumber{1} \acmArticle{1} \acmMonth{1} \acmPrice{15.00}\acmDOI{10.1145/3362063}

\begin{document}
\sloppy
%
\title{Differentiating Population Spatial Behavior using Representative Features of Geospatial Mobility (ReFGeM)}

%


\author{Rui Zhang}
\affiliation{%
  \institution{University of Saskatchewan}
  \city{Saskatoon}
  \state{Saskatchewan}
  \country{Canada}}
\email{zhang.rui@usask.ca}

\author{Kevin G. Stanley}
\affiliation{%
  \institution{University of Saskatchewan}
  \city{Saskatoon}
  \state{Saskatchewan}
  \country{Canada}}
\email{kevin.stanley@usask.ca}

\author{Daniel Fuller}
\affiliation{%
  \institution{Memorial University}
  \city{St. Johns}
  \state{Newfoundland and Labrador}
  \country{Canada}}
\email{dfuller@mun.ca}

\author{Scott Bell}
\affiliation{%
  \institution{University of Saskatchewan}
  \city{Saskatoon}
  \state{Saskatchewan}
  \country{Canada}}
\email{scott.bell@usask.ca}

%
\renewcommand{\shortauthors}{Zhang and Stanley, et al.}

%
\begin{abstract}
Understanding how humans use and consume space by comparing stratified groups, either through observation or controlled study, is key to designing better spaces, cities, and policies. GPS data traces provide detailed movement patterns of individuals but can be difficult to interpret due to the scale and scope of the data collected. For actionable insights, GPS traces are usually reduced to one or more features which express the spatial phenomenon of interest. However, it is not always clear which spatial features should be employed, and substantial effort can be invested into designing features which may or may not provide insight. In this paper we present an alternative approach: a standardized feature set with actionable interpretations that can be efficiently run against many datasets. We show that these features can distinguish between disparate human mobility patterns, although no single feature can distinguish them alone.
\end{abstract}

%
%




%

%

%
\maketitle

\section{Introduction}
Understanding movements and traces of human and animals has a notable history~\cite{miller2019towards,long2013review,wang2011human,cao2009gps,wang2019urban,zheng2015trajectory,barbosa2018human}. Spatial behavior underlies disciplines as diverse as civil engineering and commerce, architecture and anthropology, urban planning and population health. How humans use and consume space drives the design of buildings and neighborhoods, and informs health and social policy. Data on, and representations of human spatial behavior drive substantial investment and design. Classically, this data was collected using surveys, diaries or direct observations~\cite{kwan2016algorithmic}. While these data were often easy to interpret, they were subject to bias~\cite{chaix2013gps}, and characterized by sometimes insufficient spatial and temporal resolution~\cite{kwan2016algorithmic}. Automated electronic measurements, in particular GPS traces, have improved the spatial and temporal fidelity of human spatial data, but at the price of interpretability. While GPS traces provide excellent temporal spatial fidelity compared to traditional methods, they engender ethical~\cite{fuller2017ethical} concerns and have technical limitations such as battery life and a signal bias towards open outdoor spaces~\cite{stanley2016opportunistic,miller2019towards}.

High measurement frequency studies of human spatial behavior can be roughly grouped into two categories: experimental and observational~\cite{chambers2017kids,chaix2014active}. In observational studies~\cite{Schoenfelder:117085,nielsen2004behavioral}, spatial behavior patterns are observed absent a manipulation, and models are developed from data as abstractions of the behavior of the population. In experimental studies, an explicit manipulation of the spatial context is either imposed by experimenters~\cite{herrera2010evaluation}, or leveraged from a natural change in the environment~\cite{stanley2016opportunistic,dill2014bicycle,kestens2019interact} to divide observations into conditions. In both cases, comparisons between stratified data, either demographically or experimentally, is a common goal of analysis. The detail of GPS traces becomes a liability when traditional statistical analysis is applied. Statistically significant differences applied directly to detailed space time traces are phenomenologically meaningless, in fact it would be shocking in most cases if they were not different, as the number of datapoints will be sufficient to establish differences for minute effect sizes. While some insight could be obtained from examining effect sizes, or employing more sophisticated corrections, ascribing meaningful interpretation to those differences would be difficult absent a representation of how traces are different phenomenologically. Meaningful aggregation of GPS data into representative measures or features can make statistical comparisons of high measurement frequency data useful by requiring that the statistical tests are conducted on clearly interpretable movement phenomenon, and not small differences in path choice or velocity.

Feature or measure design is a fraught process. Features must preserve phenomenologically meaningful differences between individuals and populations, while substantially reducing the volume of data. Necessarily, features amplify certain spatial phenomenon at the expense of others. For example, a classic convex hull representation of activity space~\cite{fan2008urban} privileges the entire area used by an individual or population at the expense of the trajectories taken through that space. This trade off is exacerbated by the Modifiable Areal Unit Problem or MAUP~\cite{grubesic2006application}. Not only do many spatial features provide different relative values between groups at different levels of spatial and temporal analysis, that spatial dependency can be difficult to interpret when analyzing a dataset. The choice of one spatial feature over another could impact the population-level differences observed in the data. Scale free or scale interpretable features which capture a variety of spatial behaviors would be welcomed by researchers and practitioners who need to understand spatial behavior.

Traditionally, features are selected by experts based on the phenomena of interest at a spatial resolution that reflects either the behavior of interest or the resolution of the data. However, the capacity to compute these features usually far outstrips the capacity to design them in the first place. A set of standardized features, calculable from GPS data, and proven to provide actionable insight and differentiation between stratifications could provide a welcome tool for the analysis of spatial data. We proposed this idea in~\cite{zhang2018feature}, but the limited features analyzed had several shortcomings: first, we did not employ the most recent scale-free mobility entropy analysis; second, we only considered a single representation of activity space; third, we only presented differentiation between three datasets; and finally, only simple statistical models were used to verify feature utility.

In this paper, we expand on the work of~\cite{zhang2018feature}. We analyze three measures of activity space which quantify the spatial range of human movement, the latest scale free mobility entropy model which measures the predictability of movements, and fractal dimension which characterize the complexity of GPS traces. A total of nine candidate features were analyzed against six datasets from four different cities featuring a combination of demographics and movement patterns. We evaluate these features both separately against the datasets using statistical techniques, and together in a Support Vector Machine (SVM) model to demonstrate their utility. We show that while no single feature can differentiate all of the datasets, a combination of features can be successfully employed to correspond with the implicit stratifications between the datasets, and that each feature adds incremental discriminatory power to the SVM.

\section{Related Work}


There are multiple ways to record human locations over time such as Radio Frequency Identification tags (RFID)~\cite{mcnamara2008media}, social media with geo-located services~\cite{rattenbury2007towards,girardin2008digital}, Automated Fare Collection Systems (AFC)~\cite{pelletier2011smart}, GSM beacons~\cite{nanni2014transportation}, Call Detail Records (CDR)~\cite{gonzalez2008understanding,isaacman2011ranges,becker2013human}, and Global Position Systems (GPS)~\cite{ashbrook2003using}. Among these, CDR and GPS are widely used. In~\cite{gonzalez2008understanding}, researchers study the trajectory of 100,000 anonymized users with Call Detail Records to understand individual human mobility patterns. The results indicated that people follow simple, reproducible patterns despite the diversity of their travel history. CDRs are capable of tracking a large number of users; however, the data is generally sparse in time and coarse in space which limits its use in characterizing human mobility~\cite{becker2013human}. Compared to CDR, GPS in general, and smartphone-based GPS in particular, has advantages in providing more varied and finer-grain sources of location information~\cite{becker2013human}, as well as additional data from other federated sensors such as battery, WiFi, and accelerometers.


Human spatial behaviour analyses have proposed many algorithmic and methodological approaches for quantifying movement data~\cite{long2013review,fillekes2019towards}. Among these algorithms, cluster methods~\cite{lee2007trajectory} and spatial field methods such as Kernel Density Estimation are applied to discover similar movement behaviors or places of interest. Parameters which directly characterize movements such as speed, moving distance, direction can be used to differentiate movement trajectory samples~\cite{torrens2011building}. Algorithms which describe path or spatial range, such as the area of convex hull~\cite{fan2008urban} or concave hull~\cite{hu2015extracting} and entropy rate of path~\cite{song2010limits,paul2018multiscale,lin2012predictability} provide overall description of spatial behaviour and are straightforward to calculate. Methods used in analyzing animal movement behaviours can be potential features for human movement data in an integrated science of movement trajectories~\cite{miller2019towards}. For example, previous works~\cite{dicke1988using,coughlin1992swimming,torrens2012extensible} have applied fractal dimension to analyze the movement traces of clownfish larvae and human movement.

Based on the analysis of existing work, we selected nine representative features from three disciplines considering their varying characterization of human spatial behaviour. We examined buffer area, and two variations of convex hull -- widely used metrics of activity space to quantify the spatial range of human movement. From information theory we examined mobility entropy rate, a measure of the predictability of movements. We employed the latest scale-free variant of mobility entropy as described by Paul \textit{et al.} in~\cite{paul2018multiscale}. From fractal geometry we employed box counting dimension~\cite{traina2010fast} to characterize the complexity of the form of GPS traces. This feature set was chosen because it represented three distinct characterizations of a trajectory: the area covered by the trajectory, the temporal complexity of the trajectory, and the spatial complexity of the trajectory. We proposed the use of fractal dimension to describe human mobility in~\cite{zhang2018feature}, but did not explore its application across a variety of datasets. While other researchers have already applied fractal dimension to analyze the traces of human movement~\cite{torrens2012extensible}, we apply it to datasets with more samples, longer time span and more detailed daily movements.


Although features from different disciplines have been widely studied, a single feature could only characterize a specific aspect of human spatial behaviour. This inspired us to build a standard feature set for spatial behaviour, focusing on the movement of individuals measured using GPS data. In our previous work~\cite{zhang2018feature}, we employed three features, i.e., convex hull, entropy rate using the Lempel-Ziv 78 (LZ) algorithm, and fractal dimension on three datasets. Our previous work employed the simplest version of convex hull, ignored the dependence of LZ-derized entropy rate on spatial-temporal resolution, and did not look at the features in combination.

\section{Background}

In the following, we use the term feature as the general designation for all measures or metrics, drawing from the image processing and data analysis literature ~\cite{kumar2014detailed,dash1997feature}. The term feature is often overloaded in the scientific literature, and its use varies across disciplines. Here we use (spatial) feature to denote a compact representation of a spatial phenomenon or path property computed mathematically from a set of location measurements. Selecting features can be a fraught process, as features must encapsulate the phenomenon in question concisely, and be plausibly independent of other measures. Building on the proposed concept of a standard feature set in \cite{zhang2018feature}, we propose to select features with the following properties.

\begin{description}
\item[Phemenonlogically Representative] Features should be clearly attributable to the phenomenon being measured, often measured as sensitivity and selectivity in the sensor and data analytic literature. For spatial data, properties of space or trajectories should be clearly represented by the features.
\item[Rigour] Features should be underpinned by mathematically rigorous descriptions and algorithms. For spatial data, the algorithms should be mathematically stable, and have reasonable and defined computational complexity.
\item[Model Utility] Features should be able to discriminate populations in statistical models or provide greater fidelity in parametric models. For spatial features, population differences in trajectories or space use must be reliably separated.
\item[Generalizability] Features should be broadly applicable across many different data sets and sources. Spatial features should only weakly depend on the semantics of the space, and to the maximum extent possible, be independent of the spatial and temporal unit of analysis.
\end{description}

Many potential spatial features exist based on the above criteria. The most selective requirement is generalizability, as many spatial features have been designed to capture particular spatial semantics (for example shopping behavior) or are susceptible to the Modifiable Areal Unit Problem. We chose to focus our initial analysis on trajectory properties, which are broadly generalizable across spatial analysis. From the possible list of trajectory properties we chose to examine features which capture the spatial and temporal complexity of paths, on the intuition that these would differentiate populations, and the area consumed/served by the path, which would provide a spatial anchor or denominator for the path complexity measures. Fractal dimension is a logical expression of spatial complexity, and should be independent of spatial resolution. As a measure of area based on the maximum extent of a trajectory, the convex hull area is independent of the spatial unit. It may be weakly susceptible to changes in temporal sampling rate if outliers which form the boundary are excluded, but should generally be robust. Trajectory entropy rate is not independent of spatial or temporal resolution, but recent work by Paul et al. \cite{paul2018multiscale} provides a means of calculating latent path properties underlying entropy rate which are resolution independent. By combining fractal dimension, scale free entropy rate and convex hull area we hypothesize that we will be able to distinguish populations described by these trajectory features, independent of the underlying trajectory semantics.

We selected features or measures from three different mathematical disciplines to describe spatial behaviour. For detailed derivations of features presented in this section, we refer the reader to the referenced prior work. In this section, we introduce the intuitive idea of how each feature characterizes spatial behaviour and summarize the mathematics behind each feature.

\subsection{Activity Space Measures}
Activity space measures attempt to describe the area consumed during daily life. Different geographic locations, such as cities with different layouts and transportation infrastructure; rural, suburban, or urban areas; or different demographics within those constituencies (for example, employed, unemployed, or students) could have different activity spaces. Three measures of activity space, primarily drawn from computational geometry, were examined.

\textbf{Convex hull}.
Convex hull or the minimum convex polygon is a simple, intuitive, and classic method to estimate area coverage. It was first proposed by researchers in ecology to describe animal home ranges~\cite{worton1987review} and has been applied to human activity spaces for describing the geographic extent of individual daily activity patterns~\cite{fan2008urban,buliung2006urban,shareck2013examining}. It calculates the smallest polygon containing all given spatial locations, with the outermost points serving as vertices~\cite{worton1987review}. Convex hull can be illustrated with a simple thought experiment. Imagine all locations are nails dug into the ground. Extend a rubber band to enclose all the nails, then release it. When the rubber band becomes tense, it reveals the shape of the convex hull.

As a fundamental problem in computational geometry, multiple algorithms have been used to construct the convex hull of a given set of points or other objects, such as the gift wrapping algorithm~\cite{chand1970algorithm}, Graham scan~\cite{graham1972efficient}, and Quickhull~\cite{barber1996quickhull}. In this study, we employ the Python class ConvexHull from scipy.spatial which implements the Quickhull algorithm~\cite{barber1996quickhull} to compute the convex hull of all GPS locations. Using all GPS locations for an individual over the course of the study assumes that the study covers all of and only those areas participants can reach in their daily lives. This assumption strongly depends on the specific definition of daily life, and can be biased by unusual events like long-distance trips.

\textbf{Convex hull of ten locations with longest dwell time}. When we compute the convex hull above, we consider all locations visited by a participant during the study. This can unnecessarily privilege outliers -- places outside routinely visited locations at the edge of the activity space. Before the development of GPS, convex hull was constructed from the points which were typical activity locations such as home, work, and other routinely visited locations~\cite{buliung2006urban}. This classic interpretation avoids the effect of outliers but also limits the complete picture of participant mobility. An alternative formulation of convex hull which balances the outliers and spatial range of participants movements computes the area circumscribed by only those locations frequently visited. Inspired by the study~\cite{shareck2013examining} which extracted a mean of 12.7 activity locations from one-week GPS tracks, we chose 10 places where a participant dwelled for the longest time to construct a convex hull. 


\textbf{Buffer area}. As discussed in ~\cite{schonfelder2003activity}, the convex hull method assumes that people make use of the continuous space they can reach. This is a simplification of human behaviour considering the limitations imposed by the built environment. Based on the notion that areas with which people are familiar are related to their actual travel through space and constrained by transport networks~\cite{golledge1999wayfinding}, network-based approaches are also widely used to describe activity space. In these methods, activity space is encoded by buffering all of an individual's trips by a (usually fixed) distance. This method is known as buffer area or daily path mobility. Trips are commonly interpolated along road networks from incomplete GPS locations.

\subsection{Entropy rate}
Among people's daily movements, there are some trips such as going to work, purchasing groceries, or visiting the gym which constitute a routine. Alternatively, there are some incidental activities, such as a barbeque at a park, which occur spontaneously and outside of routine behaviors. The predictability or regularity of spatial behavior can differentiate individuals or populations~\cite{vanhoof2018comparing}.

 As a fundamental metric for measuring the degree of predictability of time series, entropy rate can be applied to quantify the uncertainty of individual trajectories as initially proposed by Song et al. ~\cite{song2010limits}. An approximation to entropy was described by researchers~\cite{torrens2011building} which indicates the likelihood of repetitive patterns appearing in a time series. The intuition underlying mobility entropy rate is straightforward: by partitioning the space into discrete, non-overlapping cells and assigning a label to each cell, a trajectory can be represented as a string of labels. Because the trajectory has been rendered as a string, string properties can be used to summarize the trajectory. One particular property, the entropy rate, provides the per symbol (location) information required to represent the string. Strings with a substantial repetition can be represented with less information than strings which are composed entirely of unique symbols~\cite{paul2018multiscale}.

 The entropy rate of a string is related to its compressibility and compression algorithms such as Lempel-Ziv78 (LZ) can been used for approximation. The LZ-derived entropy rate $H$ of string $S$ is given by

\begin{equation}\label{eq:LzH}
H=(\frac{1}{L}\sum\limits_{i=0}^{L-1}\Lambda_i)^{-1}\ln{L}
\end{equation}

as $L\rightarrow\infty$, where $L$ is the length of string $S$, $i$ is the index of character in $S$ and $\Lambda_i$ is the length of the minimum substring starting at $i$ which has not been observed in the substring before index $i$. However, Osgood \textit{et al.}~\cite{osgood2016theoretical} noted the LZ-derived entropy rate depends on the spatio-temporal resolution of the initial discretization of the path. To address this issue, Paul \textit{et al.}~\cite{paul2018multiscale} proposed the following model to separate the dependent (path property) and independent (measurement properties) of the path by assuming that a path could be represented by the apparent velocity and dwell time of the agents in each cell. The sampling dependent entropy rate can then be expressed as

\begin{equation}\label{eq:spatial_temporal_entropy_rate}
    H(d,T) =(d^{2}\frac{C_{1}}{4T^{2}L}+\frac{C_{2}}{4T^{2}L} +2d\frac{C_{3}}{4T^{2}L}+d\frac{C_{4}}{TL}+\frac{C_{5}}{TL})^{-1}\log L
\end{equation}


where $C_{1}=\sum_{i=1}^{n}\frac{1}{{v_{i}^{*}}^{2}}$, $C_{2}=\sum_{i=1}^{n}{t_{d_i}^{2}}$, $C_{3}=\sum_{i=1}^{n}\frac{t_{d_i}}{v_{i}^{*}}$, $C_{4}=\sum_{i=1}^{n}\frac{1}{v_{i}^{*}}$, and $C_{5}=\sum_{i=1}^{n}t_{d_i}$. $v_{i}^{*}$ is the apparent velocity across the $i^{th}$ cell and $t_{d_i}$ is the total dwell time within the $i^{th}$ cell with side length $d$. While this formulation provides a mathematically elegant decomposition of entropy rate dependence on scale and mobility variables, the marginal dwell times and apparent velocities are, by definition, not observable from the location string. To estimate the values of the marginal path properties, we follow the technique described in~\cite{paul2018multiscale} and calculate entropy rate $H$ for a number of  downsampled cell sizes $d$ and sampling rates $T$ and determine a best fit through those points according to Equation~\ref{eq:spatial_temporal_entropy_rate}, where the fit constants are the marginal path properties. Finally, we used the five constant terms in Equation~\ref{eq:spatial_temporal_entropy_rate} as features of GPS traces.

\begin{figure}[h]
  \centering
  \includegraphics[scale=0.67]{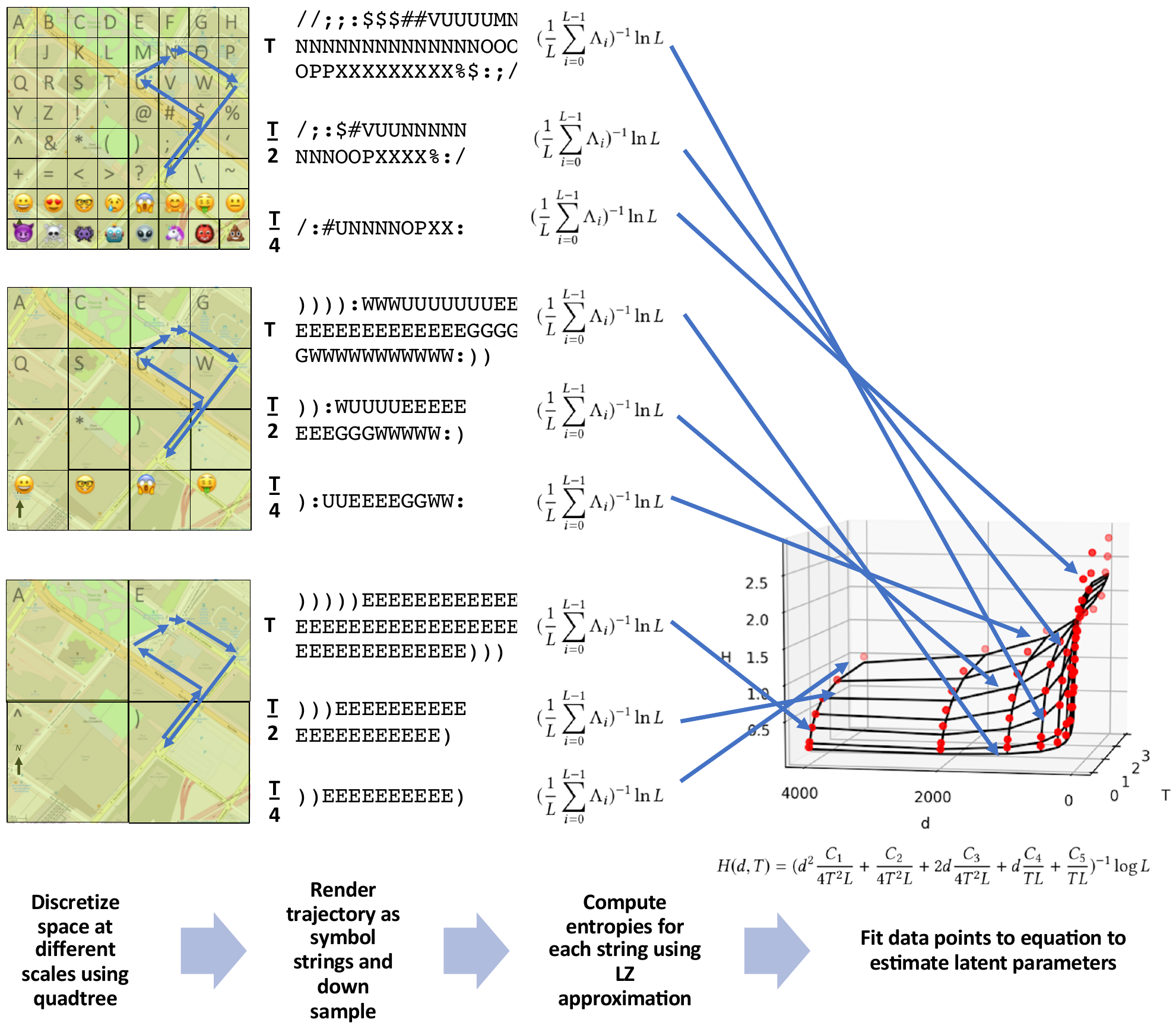}
  \caption{Schematic view of data flow for calculation of latent entropy rate variables showing three levels of spatial resolution (squares) and three levels of temporal sampling (symbol strings). Each spatio-temporal resolution becomes a point in a non linear fit of latent variables. Background maps from OpenStreetView.org.}
  \label{fig:scalefree_ent}
\end{figure}

An illustration of the data flow for computing $C1$ to $C5$ from Paul \textit{et al.}~\cite{paul2018multiscale} is shown in Fig. \ref{fig:scalefree_ent}. Space is rendered as a string through labeled discretization, in the figure represented as letters, text characters or emojis. A stylized path is represented by a sequence of arrows representing movements and dwells, starting and ending in the cell represented by the `/' symbol and dwelling for a substantial time in the cell represented by the letter `N' in the top-most discretization. Strings are down sampled regularly, using a quad-tree for space, and periodic sampling for time, represented as the three gridded maps (first column). The resulting emitted character strings corresponding to the cell visits of the hypothetical path at full, half and one quarter of the temporal sampling rate are shown in the second column. As the spatial and temporal sampling rate change, the corresponding string of visited locations represented by the characters changes as well. The corresponding entropy rates are calculated by performing LZ compression (3rd column). Each $(H,d,t)$ point is then plotted on a grid, and fit to equation (\ref{eq:spatial_temporal_entropy_rate}) using non-linear regression. $C1$ - $C5$ are the fit coefficients from (\ref{eq:spatial_temporal_entropy_rate}), and represent the path properties which give rise to entropy, independent of the scale of measurement.

\subsection{Fractal dimension}

\begin{figure}[h]
  \centering
  \includegraphics[width=\linewidth]{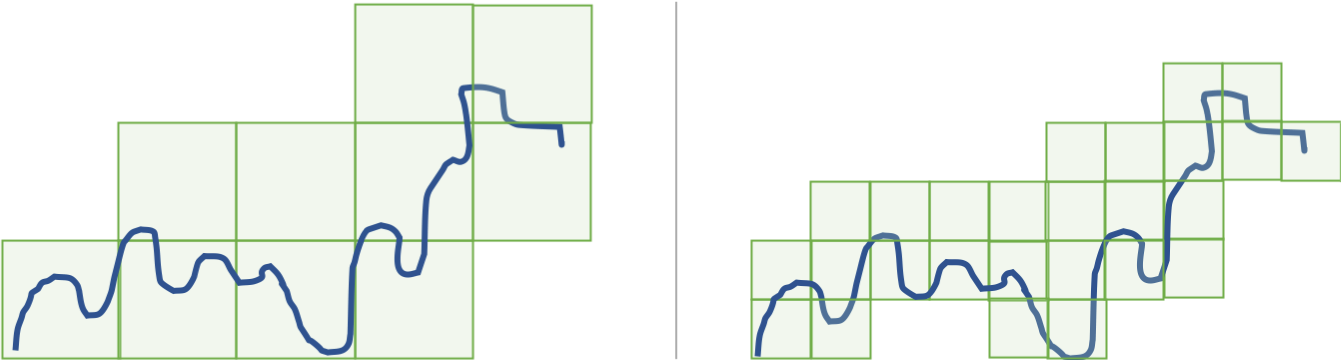}
  \caption{Illustration of box-counting dimension. A) we need 10 squares with a side length of 2 to cover the curve, B) we need 26 squares with a side length of 1 to cover the same curve.}
  \label{fig:box_conting}
\end{figure}

Fractal dimension encodes the (often fractional) number of degrees of freedom required to describe a particular curve or dataset. Simple forms can often be represented in lower dimensional traces than more complex ones. While entropy rate encodes the time complexity (particularly with respect to repetition) of trajectories, fractal dimension encodes the spatial complexity of a trajectory. The more space a subject consumes, and the more complex the paths taken to consume the space, the larger the fractal dimension. In principle, the fractal dimension of a trajectory encoded as sequence of latitude and longitude should always be less than 2, because a two dimensional surface can encode all possible trajectories. In practice, the fractal dimension of paths is typically greater than one because they can be recast mathematically as the distance along an arbitrarily long and complex curve. Given the curve, any position along it can be expressed as its distance from the start. In Euclidean geometric space, the topological dimension of these traces all equal to 1~\cite{lopes2009fractal}. In fractal geometry, the fractal dimension of a straight line equals to 1 while that of a Hexaflake equals to 1.7712~\cite{lai2012self}.
In this study, we employ the canonical box-counting method to estimate the fractal dimension. In the box-counting algorithm, space is divided into hypercubes, or in two dimensions, squares, and the number of boxes containing data is counted. The idea of box-counting algorithm is illustrated in Fig.~\ref{fig:box_conting}. Boxes are recursively sub-divided, and the number of boxes containing data at each - increasingly smaller - characteristic length is recorded. The box-counting dimension $dim_{box}$ is defined as:

\begin{equation}
dim_{box}= \underset{\epsilon\rightarrow0}{\lim}\frac{\log{N(\epsilon)}}{\log(\frac{1}{\epsilon})},
\end{equation}

where $\epsilon$ is the side length of box and $N(\epsilon)$ is the number of boxes containing data. The more constrained the curve, the smaller proportion of boxes at each level are required to represent it, and the lower the box-counting dimension.

\begin{figure}[h]
  \centering
  \includegraphics[scale=0.67]{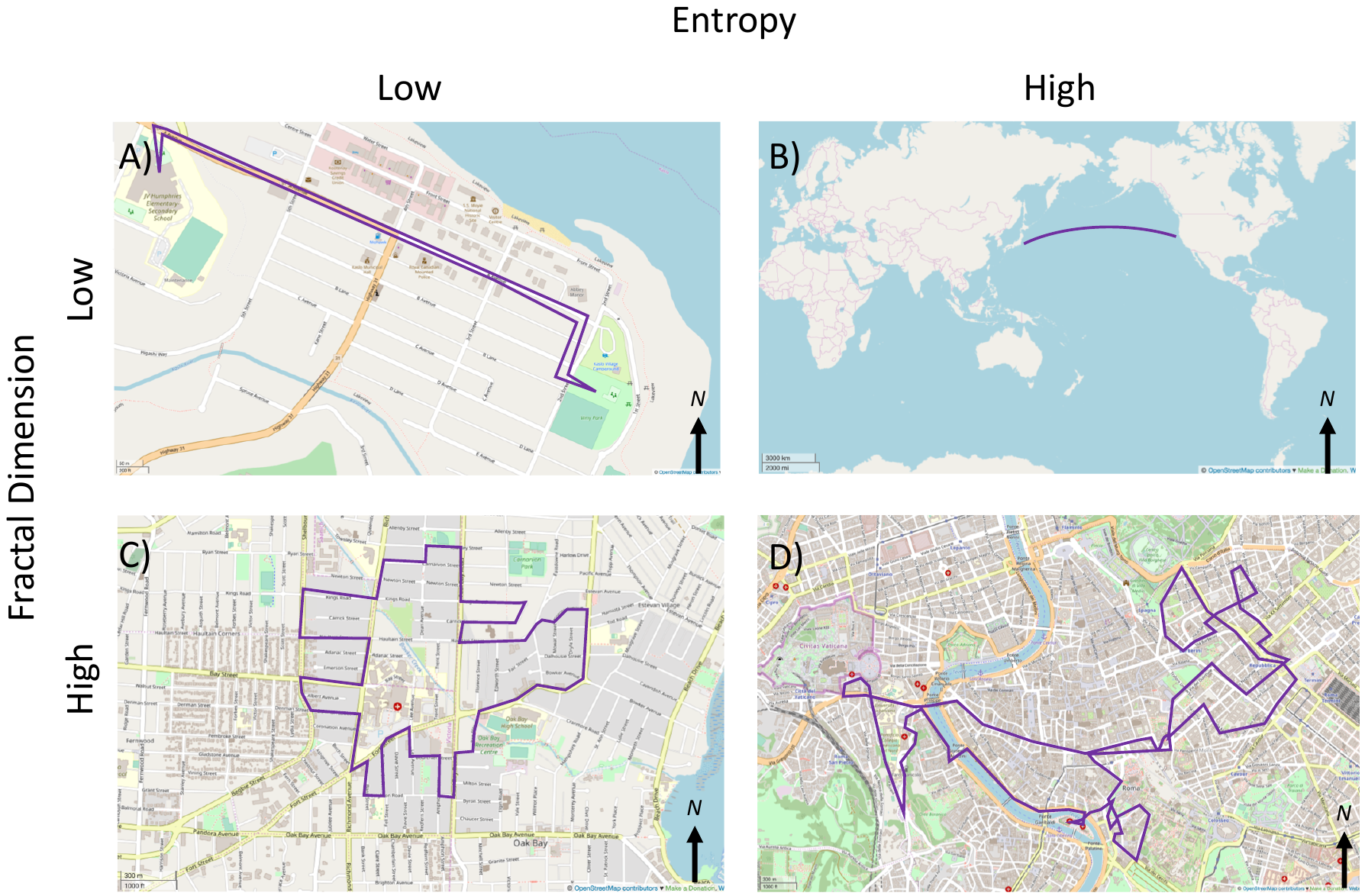}
  \caption{Schematic illustration of paths which may exhibit differences in fractal dimension or entropy. A) A rural child, B) A cargo ship, C) A suburban school bus, D) A taxi. Maps from OpenStreetMaps.org}
  \label{fig:ent_vs_fd}
\end{figure}

Entropy rate measures the spatio-temporal complexity of a trajectory, while fractal dimension measures the spatial complexity. These can often be intertwined in real trajectories, one might expect low fractal dimension paths to have low entropies, and high entropy paths to exhibit high fractal dimension. Fig. \ref{fig:ent_vs_fd} provides a schematic illustration of paths which might exhibit low or high fractal dimension or entropy rate. Each example is chosen to be a canonical example, although the cargo ship and school bus are hypothetical examples as we do not have data to verify their correctness. An elementary school-aged child living in a small town with easy walkable access to school and recreation would have a simple near linear path (low fractal dimension) and a high degree of regularity characterized by long dwells at home and school (low entropy). A cargo ship transiting the Pacific Ocean would have a simple path characterized by the great arc (low fractal dimension) but would be consistently moving to new locations, and therefore have high entropy, as calculated by the LZ approximation. A suburban school bus, which has to take a tortuous route through cul-de-sacs would have a high fractal dimension, but over many days, a low entropy as the circuit is identically repeated over and over. A taxi traveling through an old European city center would have a complex path, and limited repetition, as their destinations are dictated by their passengers.

\section{Experimental Setup}
In this paper, we explored how features from different mathematical disciplines characterize human spatial behaviour. We evaluated these features over six datasets characterized by known demographic and geographic differences, and evaluated the discrimination of each feature statistically and as input to a Support Vector Machine.

\subsection{Dataset Description}
\begin{figure}[h]
  \centering
  \includegraphics[scale=0.6]{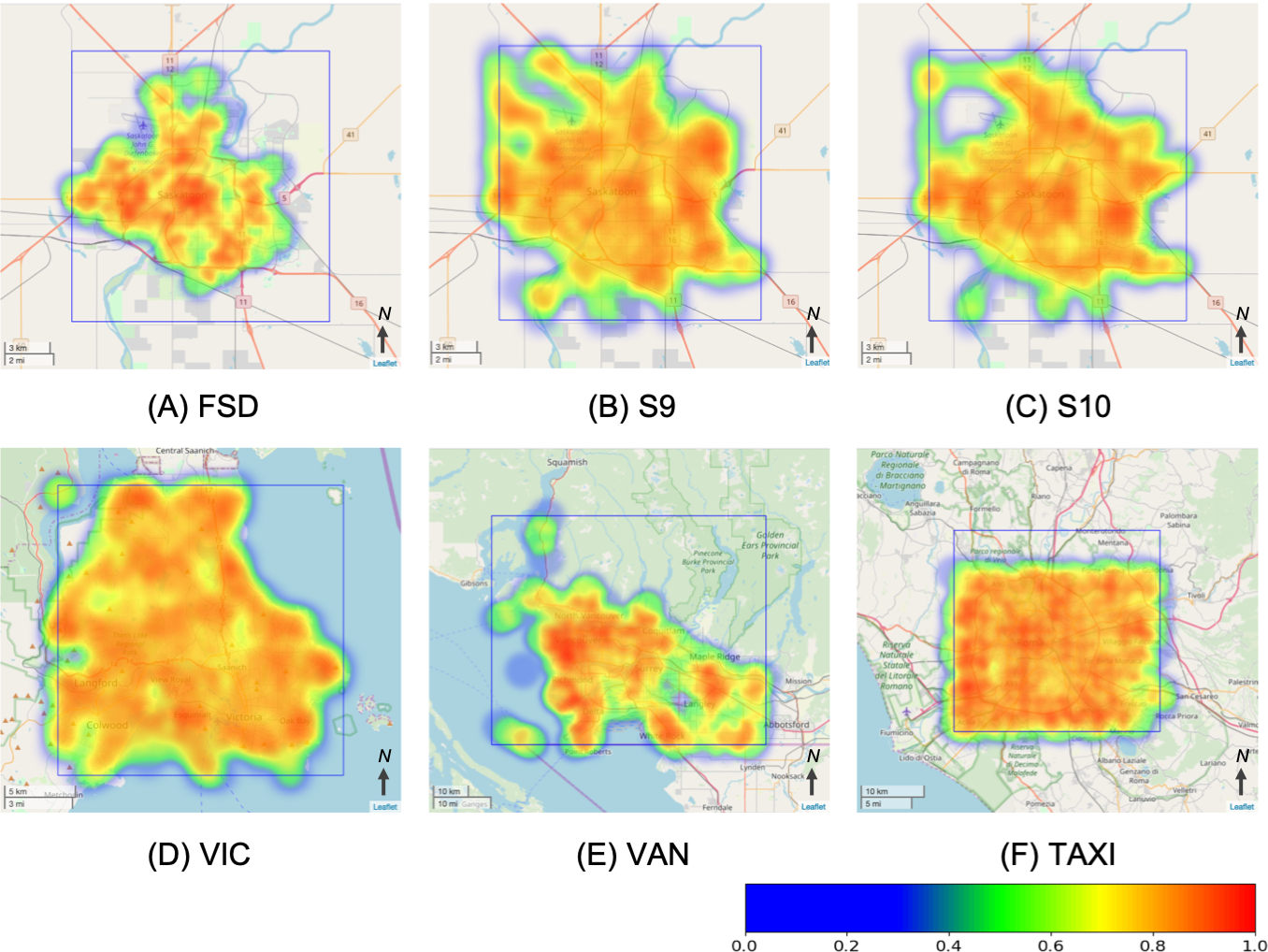}
  \caption{Heatmaps of filtered GPS records of each dataset. The blue rectangle in each heatmap depicts the defined bounding box of each city. The area within the bounding box of Saskatoon, Victoria, Vancouver, and Rome are 17 km $\times$ 17 km, 20 km $\times$ 20 km, 75 km $\times$ 63 km, and 33 km $\times$ 30 km, respectively. Scale of the map which are [FSD: 3 km, S9: 3 km, S10: 3 km, VIC: 5 km, VAN: 10 km, TAXI: 10 km] is shown at the left bottom of each each heatmap. The side length of grid for aggregating GPS records is 250 m. There are more records in the red area than the green area.}
  \label{fig:dataset_heatmap}
\end{figure}

We employed six previously collected datasets. Three of the datasets were collected from the same city, but were distinguished demographically or seasonally. The general information and demographic characteristics of each dataset are summarized in Table~\ref{tab:dataset_information} and Table~\ref{tab:dataset_demographics}. Table~\ref{tab:dataset_demographics} only summarizes demographic information of valid participants after preprocessing. The demographic information of TAXI dataset is not available. Two datasets were collected as part of the INTERACT study~\cite{kestens2019interact} in different cities. The first five datasets in Table~\ref{tab:dataset_information} all contain data covering the daily lives of participants, collected using Ethica~\cite{Ethica} or its predecessor iEpi~\cite{hashemian2012iepi} smartphone app. In all these datasets, additional sensor modalities (for example accelerometer, gyroscope and WiFi traces) were also collected, but only the GPS traces and battery data were used in this study. Battery data is used to determine data quality. If the phone is on, and Ethica is running, then battery data will be recorded, providing a more reliable measure of data quality than GPS where signals can be obscured by the built environment, but the phone is still actively recording. The final dataset was sourced from a public repository and follows taxicabs~\cite{roma-taxi-20140717} rather than individuals. The number of participants, duration and records before and after filtering can be found in Table~\ref{tab:dataset_information}. We also show the heatmap of the filtered GPS records of each dataset in Fig~\ref{fig:dataset_heatmap} for an overall view of all datasets.

Three datasets were collected from the city of Saskatoon, Saskatchewan, Canada. The food security dataset (FSD) was collected as part of a pilot study investigating novel methods for collecting data on how low-income individuals access food. Seventeen low-income families (sometimes with multiple participants) from Saskatoon were involved in the study over a three month period from April to August in 2016~\cite{osemwegie2016scalable}. The Saskatchewan Human Ethology Datasets (SHEDs) are a collection of pilot projects and technical trials related to the development of iEpi, now Ethica, and associated post-processing and methodological outcomes~\cite{knowles2014field,stanley2016opportunistic}. SHED datasets are exclusively collected from populations at University of Saskatchewan at Saskatoon, Canada. The SHED9 (S9) dataset was collected between October 28, 2016 and December 9, 2016, where 87 students including both undergraduate and graduate students but weighted towards undergraduates were observed. These participants were part of a social science student study pool. The SHED10 (S10) dataset was a similar study to S9, where 107 university students drawn from the same social science student study pool participated between February 7, 2017 and March 7, 2017. Because all three datasets were drawn from the same city, we expect them to exhibit similar spatial scales, but not activity space. FSD is distinct from S9 and S10 demographically. S9 is similar to S10, with the only notable difference being time of year (Fall versus Winter). S9 and S10 were chosen as a control. We expect them to not be discriminatable for most metrics. If metrics distinguish everything, including those things that should be similar, they may be sufficiently sensitive, but insufficiently selective.

The Vancouver (VAN) and Victoria (VIC) datasets were collected as part of the INTERACT study~\cite{kestens2019interact}. INTERACT is a five year, four site, three wave study investigating how changes in urban environments impact health. The Victoria (VIC) dataset is composed of 166 participants who are over 18 years of age and cycle at least once per month in the Greater Victoria area of British Columbia, Canada. It studies the effect of the implementation of Victoria’s All Ages and Abilities (AAA) Bike Network. The Vancouver (VAN) dataset was collected in the Greater Vancouver Area which is the third-largest metropolitan area in Canada. It is designed to reveal how the development of Vancouver's Arbutus Greenway impacts physical activity, social participation, and well-being of nearby residents. A preliminary cohort of 64 participants who are 18 years of age or older and live within 3 km of the Arbutus Greenway were recruited between May 2018 and August 2018. In both the VIC and VAN studies, Ethica smartphone traces were recorded over a one month period. The VIC dataset is distinct in geographic location, as Victoria is a coastal city approximately the same size of city area as Saskatoon, but demographically different, as participants are self-identified cycle commuters. We expect Victoria to be similar to Saskatoon in some measures of activity space, but distinct in measures which enhance temporal differences in trajectories. Vancouver is a large metropolitan area, and is expected to be distinct from all other datasets across all measures.

The Taxi dataset~\cite{roma-taxi-20140717,amici2014performance} was collected with an Android OS tablet device running an app that updated the current GPS position towards a server every 7 seconds, and is available online. It contains mobility traces of 316 taxi cabs from February 1 to March 2 2014 in Rome, Italy. The TAXI dataset is distinct in many ways from the other datasets: because it tracks taxis, not people, it is expected to have irregular trajectories, ill-defined activity spaces, and short dwell times. Because the taxis are from Rome, the trajectories should be distinct from the Canadian cities in the other datasets.

Because of conditions on our IRB ethics approval, only one of the datasets we used is available to the public (TAXI). Researchers who wish to use the SHED datasets can request to do so, but must go through a joint ethics review at both their institution and our institution. The INTERACT datasets are currently embargoed but will likely be made available under similar restrictions to the SHED data after 2021.


\setlength{\tabcolsep}{1.5pt}
\begin{table}[!ht]
\centering
\small
\caption{Datasets information}
\label{tab:dataset_information}
\begin{tabular}{|l|l|l|l|l|l|l|p{2cm}|p{2cm}|}
\hline
Dataset& City & Duration & P (\#) & P* (\#) & DC rates & Battery (\#) & GPS (\#) & GPS* (\#) \\
\hline
FSD   & Saskatoon & 90 days& 20 & 10 & 8 mins & 587722 & 1467785 & 1014949\\
S9    & Saskatoon & 40 days& 87 & 59 & 5 mins & 644367 & 15878570 & 12218667\\
S10   & Saskatoon & 29 days& 107 & 40 & 5 mins & 353840 & 8592409 & 5469949\\
VIC   & Victoria  & 30 days & 166 & 71 & 5 mins & 1024509 & 10108481 & 6606518\\
VAN   & Vancouver & 30 days& 64 & 28 & 5 mins & 341934 & 1411992 & 520437\\
TAXI  & Rome      & 30 days& 316 & 282 & 7 secs & N/A & 21817851 & 19118793\\
\hline
\multicolumn{9}{l}{P: all participants, P*: accepted participants, DC: duty cycle, GPS (\#): count of raw GPS records}\\
\multicolumn{9}{l}{GPS* (\#): count of GPS records after filtering.}
\end{tabular}

\end{table}

\begin{table}[!ht]
\centering
\caption{Demographic characteristics of each dataset (valid participants only)}
\label{tab:dataset_demographics}
\begin{tabular}{|l|l|p{1.5cm}|p{1.5cm}|p{1.5cm}|p{1.5cm}|p{1.5cm}|}
\hline
\multicolumn{2}{|l|}{\textbf{Variable}}   & FSD  & S9  & S10  & VIC & VAN  \\ \hline
\textbf{Age}            & Min    & 24          & 18         & 18          & 23          & 36          \\
                        & Max    & 63          & 37         & 38          & 67          & 76          \\
                        & Mean   & 35.5        & 25.1       & 26.1        & 42.3        & 58.1        \\
                        & Std    & 13.9        & 4.8        & 5.3         & 11.9        & 10.6        \\ \hline
\textbf{Gender}         & Female & 8           & 37         & 25          & 31          & 16          \\
                        & Male   & 2           & 22         & 15          & 38          & 12          \\
                        & Other  & 0           & 0          & 0           & 2           & 0           \\ \hline
\end{tabular}
\end{table}

\subsection{Data Preprocessing}
We applied a three-step preprocessing on each dataset for further feature analysis. The preprocessing stage includes data filtering, transformation and aggregation.

\textbf{Filtering}. We first removed unreliable erroneous GPS locations with an accuracy poorer than 100 meters. We also removed locations falling outside of the defined bounding box of a city because  we are interested in the spatial behaviour within each city area. City bounding boxes employed are listed in Table~\ref{tab:city_bounding_box}. We excluded participants with less than half of the maximum battery duty cycles (phone is on and iEpi or Ethica are running) and a quarter of the maximum possible GPS duty cycles.

\begin{table}[!ht]
\centering
\caption{Range of each city\textquotesingle s bounding box}
\label{tab:city_bounding_box}
\begin{tabular}{|p{2cm}|p{4cm}|p{4cm}|}
\hline
City & Range of latitude  & Range of longitude  \\ \hline
Saskatoon & (52.058367, 52.214608) & (-106.764914, -106.522253)   \\ \hline
Victoria  & (48.391892, 48.57277) & (-123.540331, -123.271128)     \\ \hline
Vancouver & (49.001407,  49.566829) & (-123.441453, -122.406227)     \\ \hline
Rome & (41.769653, 42.052603) & (12.341707, 12.730937)   \\ \hline
\end{tabular}
\end{table}
\textbf{Transformation}. The GPS records in all datasets were recorded as latitude and longitude and were converted to UTM (Universal Transverse Mercator) coordinates (easting, northing) using pyproj 1.9.5.1. Following the entropy rate calculation in~\cite{paul2018multiscale}, continuous city space was discretized into square bins with a side length 4 km, then down-sampled using a quad-tree decomposition to a minimum cell size of 15.625 m. The index $(x, y)$ of each bin was calculated using

\begin{equation}
 \begin{aligned}
    x = \frac{easting - easting_{min}}{bin\_size},\\
    y = \frac{northing - northing_{min}}{bin\_size}
 \end{aligned}
\end{equation}

where $easting_{min}$ and $northing_{min}$ are the lower bound of the corresponding bounding box of each city.

Unless otherwise noted, interpolation is not employed when computing features based on GPS traces. For entropy, which requires a contiguous string of locations to calculate outcomes, gaps in recording were ignored and location strings were constructed ignoring time gaps as in~\cite{song2010limits,paul2018multiscale,osgood2016theoretical}. Features were calculated for all GPS records available over the course of each study for each participant. In the results, the primary unit of analysis is the participants GPS trace over the entire study period. While these features also permit stratification through time or space (for example, before and after an intervention, or near work or near home), only between dataset analyses using individual participant traces are presented.

\textbf{Aggregation}. In all datasets, multiple GPS timepoints were recorded in a single duty cycle. To ensure that we had as regularly-spaced time steps as possible we followed the strategy in~\cite{paul2018multiscale}, representing the location of each duty cycle with the first record in that duty cycle.

\subsection{Configuration of feature calculations}
With the above preprocessing, three types of representations of locations were available for analysis: geographical coordinates, UTM coordinates, and indices $(x, y)$ corresponding to the discretized space. Because distance calculations and binning calculations are simpler in UTM coordinates, we employed UTM coordinates instead of latitude and longitude to calculate these features. Additionally, once we had calculated UTM coordinates for one feature, it was simplest to employ them for all features. UTM coordinates and grid index were applied to different features following the literature and considering computational complexity~\cite{sherman2005suite,paul2018multiscale,paul2018mobility}.

\textbf{Convex hull}. Considering that specific locations encoded by grid index are required to calculate the convex hull of ten locations with longest dwell time, we also employed grid index for the convex hull calculation for consistency. Because the grid index under base bin size (15.625 m) has been calculated in preprocessing, we directly applied the Python class \texttt{ConvexHull} from scipy.spatial which implements the Quickhull algorithm to the grid index of all records to get the area of convex hull.

\textbf{Convex hull of ten locations with longest dwell time}. To extract the ten locations with longest dwell time from GPS records, we used the index of each grid cell to aggregate locations. The side length of the grid is 250 meters instead of 15.625 meters in order to cluster movements within a place. Following the studies that often use size ranges between 200 meters and 500 meters~\cite{montoliu2010discovering,thierry2013detecting}, we chose 250 meters as the maximum distance to be considered a dwelling location. A dwell starts when consecutive GPS records fall into the same grid, and ends when the GPS location is outside the grid. Every dwell is summed to get the total dwell time in each grid cell. The 10 places with longest dwell time were extracted and the same Convex Hull class was employed to calculate the area of convex hull.

\textbf{Buffer area}. As we are only interested in the area of the buffered trip, we used the straight line between two locations as a simplification of real road network. Following~\cite{hirsch2014generating}, we took 200 meters as the buffer distance. UTM coordinates were used in calculating buffer area. Each two consecutive locations were connected with a straight line and treated as a segment of the whole path. Each segment was buffered using the \texttt{buffer} function in Python package shapely 1.6.4.post2. Finally, all buffers were combined using the \texttt{cascaded\_union} function in the same package to get the overall buffer area.

\textbf{Entropy rate}. To approximate the constant terms $C_{1}$ to $C_{5}$, we followed the technique proposed in~\cite{paul2018multiscale}. First, we calculated the entropy rate $lzH$ over pairs of $(T,d)$ of the sequence of GPS locations represented as $(x,y)$ according to Equation~\ref{eq:LzH}. The down-sampling intervals $T$ of different datasets are shown in Table~\ref{tab:downsampling intervals}. The range of spatial quantization $d$ is [15.625 m, 31.25 m, 62.5 m, 125 m, 250 m, 500 m, 1 km, 2 km, 4 km] following~\cite{paul2018multiscale}. The resulting set of $(T, d, L, lzH)$ was analyzed using Eureqa~\cite{dubvcakova2011eureqa} to get the constant terms $C_{1}$ to $C_{5}$ from (\ref{eq:spatial_temporal_entropy_rate})for each participant.

\begin{table}[!ht]
\centering
\caption{Down-sampling intervals of different datasets}
\label{tab:downsampling intervals}
\begin{tabular}{|l|l|}
\hline
Dataset & Down-sampling intervals                              \\ \hline
FSD     & 8 min, 40 min, 80 min, 2 hr, 4 hr, 8 hr              \\ \hline
S9      & 5 min, 10 min, 30 min, 1 hr, 2 hr, 4 hr, 8 hr        \\ \hline
S10     & 5 min, 10 min, 30 min, 1 hr, 2 hr, 4 hr, 8 hr        \\ \hline
VIC     & 5 min, 10 min, 30 min, 1 hr, 2 hr, 4 hr, 8 hr        \\ \hline
VAN     & 5 min, 10 min, 30 min, 1 hr, 2 hr, 4 hr, 8 hr        \\ \hline
TAXI    & 1 min, 5 min, 10 min, 30 min, 1 hr, 2 hr, 4 hr, 8 hr \\
\hline
\end{tabular}
\end{table}

\textbf{Fractal dimension}. We employed Paul\textquotesingle s implementation~\cite{paul2018mobility} of box-counting dimension which builds a n-Dimentional Tree~\cite{traina2010fast}. Following this implementation, we used the grid index to aggregate GPS records to duty cycles as described in Preprocessing. The implementation can be rendered algorithmically as
\begin{enumerate}
    \item Remove repeated locations to avoid endless loops in the process of building n-Dimensional Tree.
    \item Employ n-Dimensional Tree to produce a set of tuples of the form
            $\left[\log\frac{1}{\epsilon},\log{N(\epsilon)}\right]$.
    \item Fit a least-square regression line for $\log{N(\epsilon)}$ versus $\log\frac{1}{\epsilon}$.
    \item Estimate the box-counting dimension as the slope of regression line.
\end{enumerate}

\textbf{Experimental environment}. All the experiments were run on a MacBook Pro with 2.4 GHz Intel Core i5 and 8GB 1600 MHz DDR3 RAM. The general processing of GPS records and activity space features calculation were done using Python 3.6.7 with PyCharm 2017.1.3. Packages included Pandas 0.23.0, Numpy 1.15.4, scikit-learn 0.20.1, scipy 1.1.0, pyproj 1.9.5.1, folium 0.7.0, geopandas 0.4.0, shapely 1.6.4.post2, and matplotlib 3.0.2. Entropy rate and box-counting dimension were calculated using Paul\textquoteright{}s implementation~\cite{tuhin2017entropyrate,paul2018mobility}. R packages ggplot2 3.1.0 and ggpubr 0.2 under R 3.5.0 were also used to draw boxplots in results. Functions aov and TukeyHSD in R were employed for one-way ANOVA test and Tukey's HSD test, respectively. The code used to implement these algorithms is publicly available and can be found at~\cite{rui2019ReFGeM}.

\subsection{Classification}
We extracted nine features in total from the GPS locations of each participant. The features were the area of convex hull (CH), the area of convex hull of ten locations with longest dwell time (CH10), the buffer area (BA), the five constant terms ($C1$ to $C5$) of entropy rate considering varying spatial and temporal resolution, and the box-counting dimension (DIM). To examine the power of combinationing features for discriminating individuals, we trained a multi-class Support Vector Machine (SVM) model. We validated the effect of each feature on the overall classification performance by adding each feature incrementally to the classifier and rerunning the classification to compare the performance.

The models and feature selection were implemented with Python package scikit-learn 0.20.1. Because we hope to know the order of importance of each feature on the classifier, we employed the \texttt{selectKBest} function to reveal the complete order of importance according to the ANOVA F-value metric. Other rankings of features are possible (for example by principle component analysis), but because this evaluation is meant to be primarily illustrative, those evaluations are left for future work. By assigning $k$ values from 1 to 9 to \texttt{selectKBest} function, features of higher importance were selected first.

In the training step of SVM classifier, we tested the classifier on unbalanced datasets as well as balanced dataset with weights: [FSD: 28, S9: 5, S10: 7, VIC: 4, VAN: 10, TAXI: 1]. The weight is inversely proportional to the number of participants in that dataset. We used 75\% of the dataset as training set and the remaining 25\% as test set. We tested different kernels for SVM including linear, polynomial, RBF, and sigmoid kernels. The linear kernel outperformed other kernels and was used in the classification model. Finally, we employed 5-fold cross validation to derive the mean accuracy score and standard deviation value for the model performance evaluation.

\section{Results}
To determine the utility of the proposed features, the ability of each feature to discriminate between each dataset in isolation was evaluated. By examining each feature in isolation, we can form hypotheses as to what phenomena it is sensitive to and selective for, and test those against our intuitions. To determine the utility of the features employed jointly, an SVM was trained to assign participants to the studies from which they originated.

\subsection{Single Feature Analysis}
\begin{figure}[h]
  \centering
  \includegraphics[width=\linewidth]{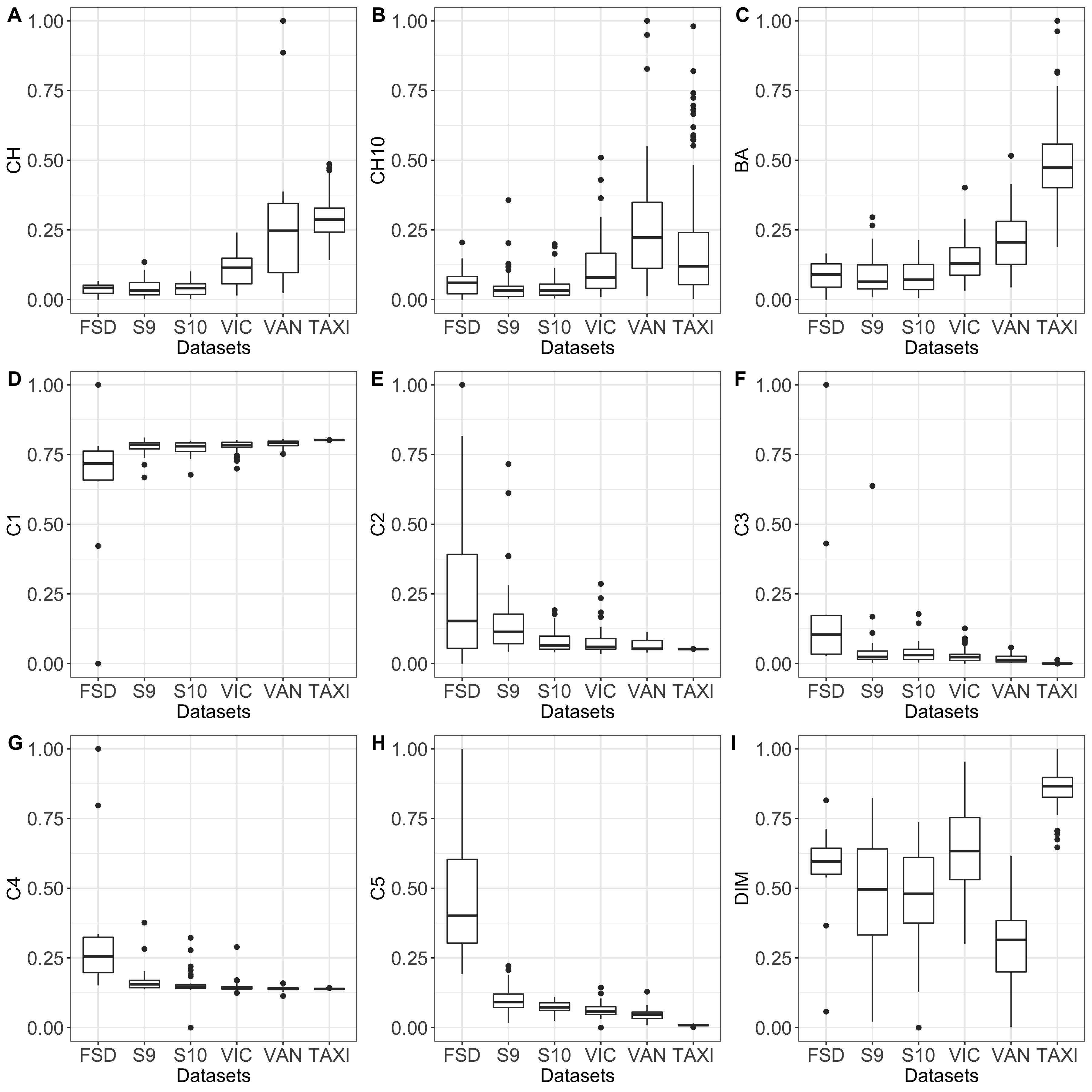}
  \caption{Distribution of each feature over all datasets: \textbf{(A)} convex hull \textbf{(B)} convex hull of ten locations with longest dwell time \textbf{(C)} buffer area \textbf{(D - H)} constant term $C1$, $C2$, $C3$, $C4$, and $C5$ of entropy rate \textbf{(I)} fractal dimension. The lower and upper hinges correspond to the first and third quartiles. The inner line within the box represents the median of the data. The upper whisker represents the range between the third quartile and the largest value but no larger than 1.5 $\times$ IQR (interquartile range). The lower whisker represents the range between the first quartile and the smallest value but at most 1.5 $\times$ IQR. Data beyond the end of the whiskers are outlying points and are plotted as solid circles. }
  \label{fig:all_features_boxplot}
\end{figure}

To demonstrate the discriminatory capability of each feature, we normalized each feature over all datasets and plotted the distribution of each metric in Fig~\ref{fig:all_features_boxplot}. Each panel in the figure represents the distribution of each measure across full records for each participant in that study; that is each participant's value for each feature are calculated for each dataset. The distribution of feature values across participants for each study are then rendered as boxes in the boxplot. Substantial differences between the relative values for each dataset are evident across the features, but no consistent pattern is evident, providing us with confidence that different features are enhancing different phenomenon in the observed spatial traces.

A two-step statistical analysis was applied to each metric. First, we used a one-way analysis of variance (ANOVA) to test the difference of distributions over all datasets, and subsequently, a Tukey's HSD test to discover the significantly different pairs of datasets. Because of the number of samples and discriminatory power of the features, the fact that the ANOVA was always highly significant is not meaningful for our analysis. We only report the results of the Tukey\textquotesingle s HSD tests, as summarized in Table \ref{tab:entropy_rate_tukeys}. Data is reported to three decimal places for readability. Significance values close to 1 are reported as >0.999 and small significance values are reported as <0.001. Taking a standard 0.05 level as significant, all significant results are rendered in bold. Several trends are clearly evident from the table. FSD and VAN are always different, separated by both geography and demographics. All other datasets are almost always different from TAXI, except for VAN, as expected as taxi patterns would be different than individuals in most circumstances. Individual differences in datasets are discussed in the following subsections.

\begin{table}[!ht]
 \caption{P-value of Tukey's HSD test on all features}
  \label{tab:entropy_rate_tukeys}
\begin{tabular}{|l| c|c|c | c|c|c|c|c | c|}
\hline
 & \multicolumn{3}{c|}{\textbf{Activity Space}} & \multicolumn{5}{c|}{\textbf{Entropy Rate}} & \multicolumn{1}{c|}{\textbf{Fractal}} \\ \hline
Dataset pairs  & CH      & CH10    & BA      & $C1$      & $C2$      & $C3$      & $C4$      & $C5$     & DIM \\\hline
FSD-S9   & >0.999 & 0.997 & >0.999  & \textbf{<0.001} & \textbf{<0.001} & \textbf{<0.001} & \textbf{<0.001} & \textbf{<0.001} & 0.446\\
FSD-S10  & >0.999 & 0.998 & >0.999  & \textbf{<0.001} & \textbf{<0.001} & \textbf{<0.001} & \textbf{<0.001} & \textbf{<0.001} & 0.289 \\
FSD-VIC  & 0.074 & 0.946 & 0.580  & \textbf{<0.001} & \textbf{<0.001} & \textbf{<0.001} & \textbf{<0.001} & \textbf{<0.001} & 0.286\\
FSD-VAN  & \textbf{<0.001} & \textbf{<0.001} & \textbf{0.013}  & \textbf{<0.001} & \textbf{<0.001} & \textbf{<0.001} & \textbf{<0.001} & \textbf{<0.001} & \textbf{<0.001}\\
FSD-TAXI & \textbf{<0.001} & 0.240 & \textbf{<0.001}  & \textbf{<0.001} & \textbf{<0.001} & \textbf{<0.001} & \textbf{<0.001} & \textbf{<0.001} & \textbf{<0.001}\\
S9-S10   & >0.999 & >0.999 & >0.999  & 0.994 & \textbf{<0.001} & >0.999 & 0.955 & \textbf{0.016} & 0.993\\
S9-VIC   & \textbf{<0.001} & 0.084 & \textbf{0.037}  & >0.999 & \textbf{<0.001} & 0.659 & 0.209 & \textbf{<0.001} & \textbf{<0.001}\\
S9-VAN   & \textbf{<0.001} & \textbf{<0.001} & \textbf{<0.001}  & 0.895 & \textbf{<0.001} & 0.370 & 0.170 & \textbf{<0.001} & \textbf{<0.001}\\
S9-TAXI  & \textbf{<0.001} & \textbf{<0.001} & \textbf{<0.001}  & \textbf{<0.001} & \textbf{<0.001} & \textbf{<0.001} & 0.002 & \textbf{<0.001} & \textbf{<0.001}\\
S10-VIC  & \textbf{<0.001} & 0.201 & 0.054  & 0.987 & >0.999 & 0.878 & 0.870 & 0.809 & \textbf{<0.001}\\
S10-VAN  & \textbf{<0.001} & \textbf{<0.001} & \textbf{<0.001}  & 0.701 & 0.904 & 0.579 & 0.673 & 0.078 & \textbf{<0.001}\\
S10-TAXI & \textbf{<0.001} & \textbf{<0.001} & \textbf{<0.001}  & \textbf{<0.001} & 0.109 & \textbf{<0.001} & 0.224 & \textbf{<0.001} & \textbf{<0.001}\\
VIC-VAN  & \textbf{<0.001} & \textbf{<0.001} & \textbf{0.036}  & 0.914 & 0.958 & 0.964 & 0.988 & 0.425 & \textbf{<0.001}\\
VIC-TAXI & \textbf{<0.001} & \textbf{0.031} & \textbf{<0.001}  & \textbf{<0.001} & 0.067 & \textbf{<0.001} & 0.853 & \textbf{<0.001} & \textbf{<0.001}\\
VAN-TAXI & 0.401 & \textbf{<0.001} & \textbf{<0.001}  & 0.465 & 0.949 & 0.462 & >0.999 &\textbf{<0.001} & \textbf{<0.001}\\
\hline
\end{tabular}
\end{table}

\subsubsection{Convex hull and its variation}

The distribution across participants by dataset of the area of standard convex hull is shown in Fig~\ref{fig:all_features_boxplot}-A. Tukey's HSD test shows that there is no significant difference between FSD, S9, S10 collected in the same city (Saskatoon), as expected. There is no significant difference between datasets TAXI and VAN. The three Saskatoon datasets (FSD, S9, and S10) are different from TAXI and VAN. FSD is not different from VIC, but  S9 and S10, both studies in Saskatoon are different from VIC. Plausible explanations for these differences include that FSD participants were low income and had their behaviour constrained by mobility challenges due to income, while the VIC samples had their behaviour constrained by the topography of a coastal city~\cite{park2017multi}. The S9 and S10 datasets differed from VIC because students may be less constrained in their activity space than low income individuals, perhaps partly due to the free bus pass program at the university~\cite{stanley2016opportunistic}.


The plot in Fig~\ref{fig:all_features_boxplot}-B shows the distribution of area of convex hull built with ten locations with longest dwell time. Although this feature is also an area of the convex hull, the Tukey results are different with the standard convex hull. VAN is different from all the other datasets, which is not surprising given its scale. TAXI is different from S9, S10, VIC, and VAN, which again is as expected as a taxi's top ten locations will be dictated by the whims of its customers more than the established geographic patterns of individuals. There is no significant differences between all Saskatoon datasets and VIC which is different from the results from standard convex hull. It shows that a tighter activity space between similar sized cities may be expected. It is somewhat surprising that TAXI and FSD are not different. There is no significant difference between any dataset collected in Saskatoon, the same result obtained from the standard convex hull.

\subsubsection{Buffer Area}
Buffer area reveals the space surrounding the locations visited by individuals during their daily movements, and is distinct from convex hull which instead describes the space circumscribed by the extents of people's movements. Fig~\ref{fig:all_features_boxplot}-C is the distribution of buffer area normalized over all datasets. According to the post hoc analysis, there is no significant difference between the Saskatoon datasets. VIC is not significantly different from FSD and S10. All the other pairs are significantly different.


\subsubsection{Entropy Rate}
\begin{figure}[h]
  \centering
  \includegraphics[scale=0.55]{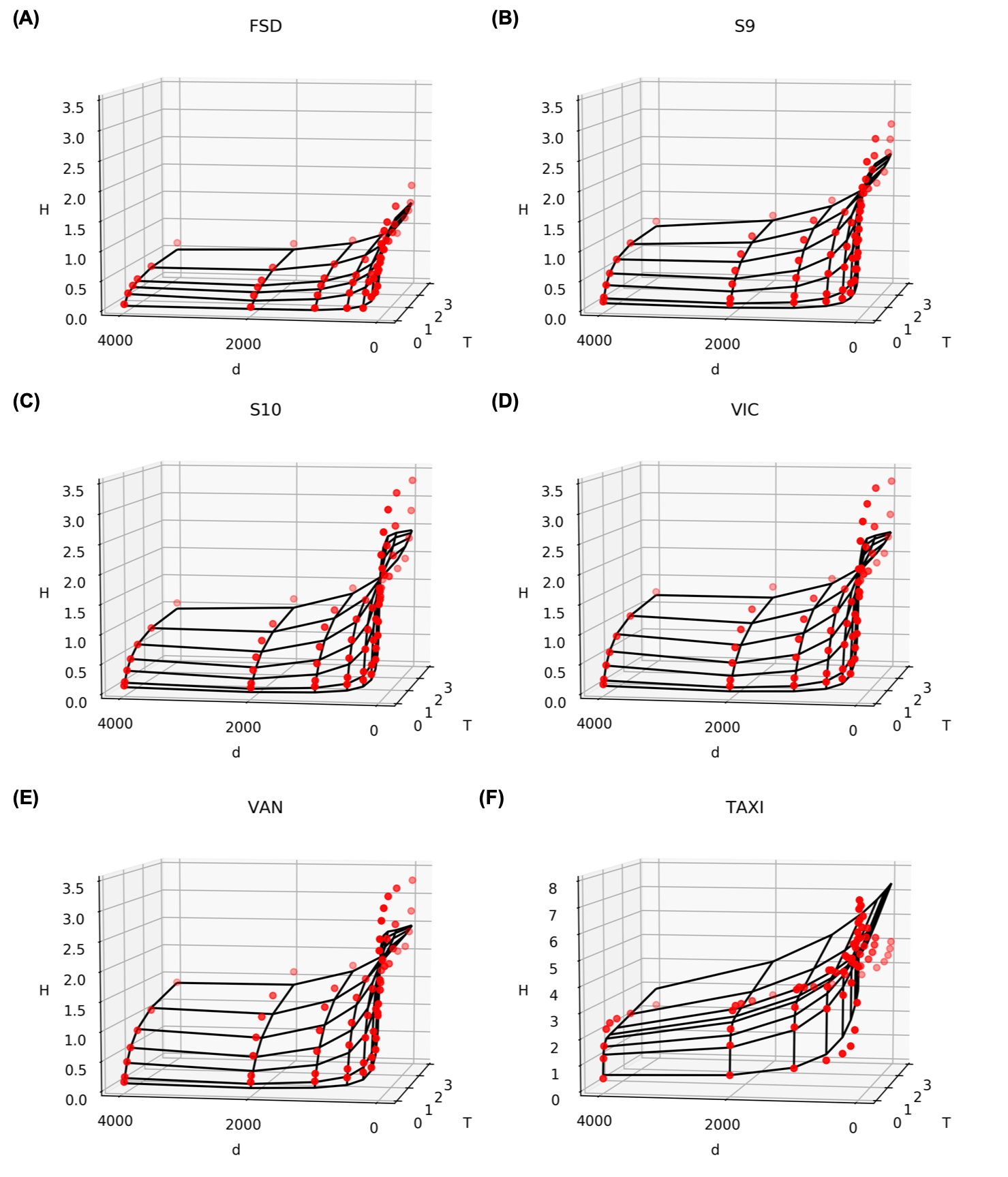}
  \caption{Entropy surface and empirical points for A) FSD, B) S9, C) S10, D) VIC, E) VAN, and F) TAXI. d is in meters, T is in $10^4$ seconds, and H is in bits. The wire frame surface is constructed with the average of entropy rate H calculated following Equation~\ref{eq:spatial_temporal_entropy_rate} with fitted constant terms. The scatter points are the average of LZ-derived entropy rate following Equation~\ref{eq:LzH} over each dataset. Taxis  exhibited the greatest entropy rate and therefore has a different scale for the z-axis.}
  \label{fig:entropy_surface_fit}
\end{figure}

The dependence of mobility entropy rate on spatial and temporal resolution $H(d,T)$ can be expressed by the five constant terms from~\cite{paul2018multiscale}. Surfaces denoting the model with fitted constant terms, and the average calculated LZ-derived entropy rate over each dataset, are shown in Fig~\ref{fig:entropy_surface_fit}. It is clear that the model proposed in~\cite{paul2018multiscale} provides an accurate description of how mobility entropy rates vary across a wide range of spatial and temporal scales. As proposed in~\cite{paul2018multiscale} these constants themselves can potentially serve as features. This is the first work we are aware of which examines the utility of these terms as features.

The distributions of $C1$ - $C5$ are plotted in Fig~\ref{fig:all_features_boxplot}-(D-H). The ANOVA test reveals a significant difference in all constant terms among datasets. In Table~\ref{tab:entropy_rate_tukeys}, the following patterns are kept for all constant terms:
\begin{enumerate}
    \item FSD is always significantly different from all the other datasets.
    \item There is not a significant difference between VAN and S10, VIC and S10, and VAN and VIC.
    \item In general, the constant term $C5$ is able to distinguish the most pairs of datasets (12/15).
\end{enumerate}

\subsubsection{Fractal Dimension}
Tukey's HSD test shows that there is a significant difference between most pairs of datasets, but no significant difference between FSD and VIC, FSD and S9, FSD and S10, and S9 and S10. Datasets from similarly sized cities appear to have the similar fractal dimension. This could be in part because a greater area provides more opportunity for more complex paths.

\subsection{Application to Machine Learning}
From the above experiments and results, we can demonstrate that no single feature can distinguish these datasets, which is not surprising. But most datasets were distinguishable by some non-overlapping set of features, with the exception of SHED9 and SHED10 which were intentionally selected to be similar. To investigate the power of using combinations of features, we employ the combinations of nine features for individual discrimination with an SVM.

Following the feature selection step using \texttt{selectKBest} function in scikit-learn package (which sorts based on F-value), the features ordered by importance are [BA, $C5$, DIM, CH, $C4$, $C2$, $C1$, $C3$, CH10]. We note that this is a single greedy ordering of features, and that other tests could generate other feature orderings. We built a sequence of SVM classifiers, increasing the number of included input features, with the goal of assigning participants to datasets. After training the SVM, the average accuracy was calculated from five-fold cross-validation. The accuracy score of five-fold cross validation is 0.36 (+/- 0.22) for a single feature (BA). When the second important feature ($C5$) was added to the feature set, the accuracy score increased to 0.72 (+/- 0.05). As expected, the accuracy increased monotonically with additional features, but with declining return. For the data examined here, most of the accuracy is gained after the first six to seven features, incorporating all three major classifications of features, the two most distinct activity space features, and entropy fit features corresponding to marginal dwell time and apparent velocity.

\begin{figure}[h]
  \centering
  \includegraphics[width=\linewidth]{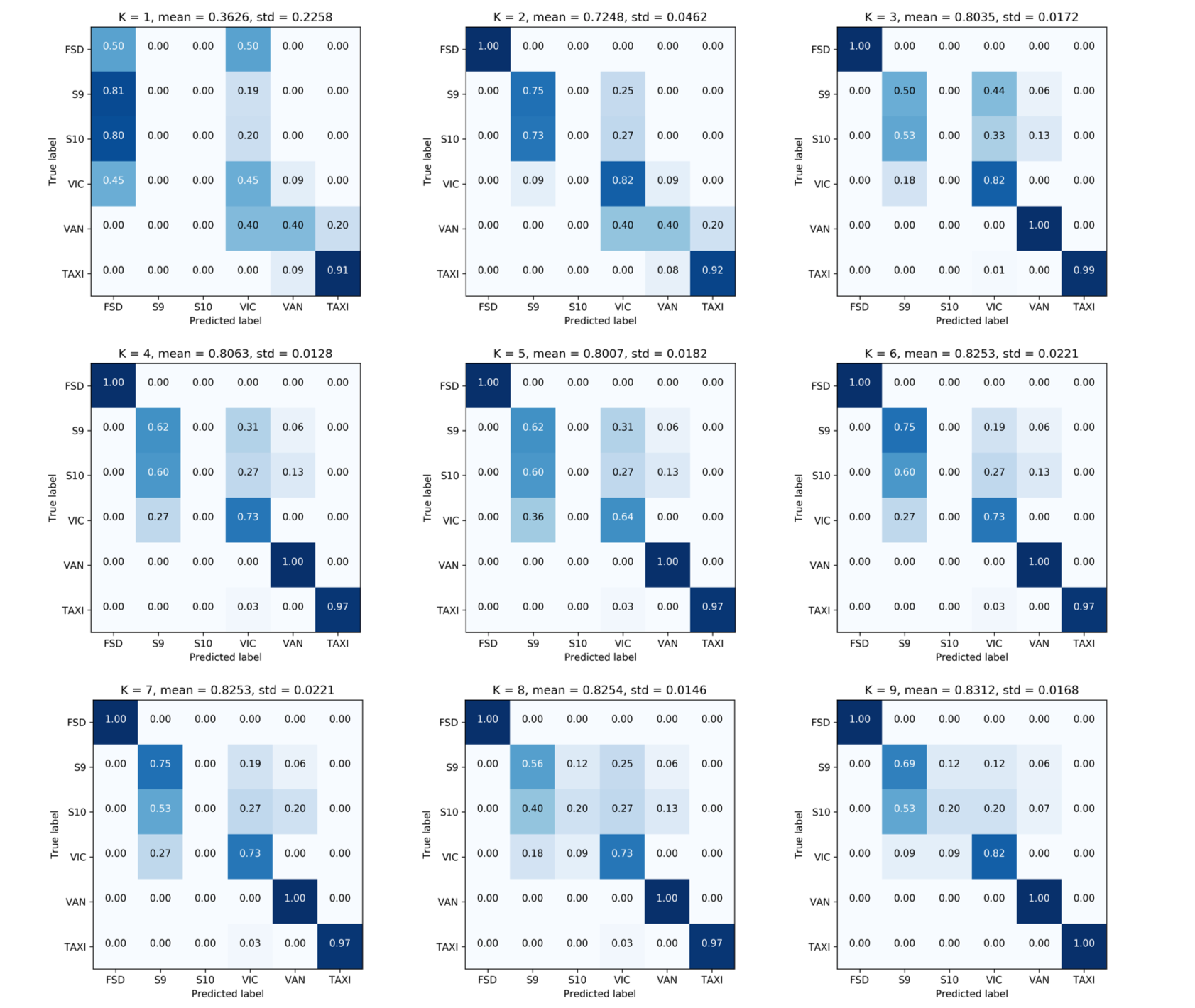}
  \caption{Confusion matrix of SVM classifier on test set. Increasing number of features were used in training SVM model. Above each confusion matrix, K denotes the count of features used in SVM, mean represents the mean accuracy of five fold cross validation on training set, and std is the standard deviation of cross validation on training set.}
  \label{fig:pred_results}
\end{figure}

The confusion matrix of all SVM classifiers are shown in Fig~\ref{fig:pred_results}. It is worth noting, that unlike the statistical tests which endeavored to differentiate populations, the SVM differentiate individuals. The model is attempting to answer the question: given an individual, which dataset are they from? This is a much more complex task. For a single feature, the classifier assigns most individuals to FSD and TAXI. The subsequent two features ($C5$, DIM), allow the near perfect assignment of participants to FSD, VAN and TAXI, with a reasonable number of true positives for VIC, but a large number of false positives, predominately arising from S9 and S10. Adding CH, $C4$ and $C2$ help to distinguish S9 from VIC, but S10 remains poorly classified, contributing a number of false positives to both S9 and VIC. $C3$ could distinguish some participants of S10 from S9 and VIC, but still in a poor condition. Additional features have marginal impact on the overall accuracy, and simply shift the false positives from S9 and S10 between each other and VIC.



\section{Discussion}

The results demonstrate the utility of using Representative Features of Geospatial Mobility (ReFGeM) to distinguish different datasets from each other. No single feature was able to statistically separate all datasets. While fractal dimension came close to being able to differentiate between all datasets, it could not distinguish between VIC and FSD, or FSD and S9 or S10. S9 and S10 were meant to be closely correlated datasets and were only distinguished by the dwell-time proportional constants in the scale free entropy calculation. This effect could have been due to the differences in student behavior during different seasons (Fall and Winter). As expected, measures corresponding to activity space were diagnostic of the city the dataset was collected in. Convex hull is a classic, but potentially limited measure while buffer area is a more recent implementation. Buffer area is similarly diagnostic to convex hull, but slightly less prominent. Fractal dimension is diagnostic of demographic differences between datasets, with taxis, cyclists and lower income Saskatoonians experiencing more fractal movement patterns than datasets demographically skewed towards students.

Scale-free entropy measures correspond to five individual constants, and are therefore the most difficult to interpret. A simple method to understand the entropy rate results are to map the constants to the value being summed over cells, $C1$ is one over the squared velocity, $C2$ is the squared dwell time, $C3$ is the quotient of dwell time and velocity, or inverse distance, $C4$ is velocity, and $C5$ is dwell time. FSD is different than all other datasets across all constants. While this may have a demographic basis, FSD is also characterized by the smallest and most sporadic per-participant GPS traces of all the datasets. The work by Paul \textit{et al.}~\cite{paul2018multiscale} assumes that traces are sufficiently dense and long to support spatial decomposition and temporal sampling with no need for interpolation. With this assumption, the LZ-Entropy approximation can converge, and the fits parameters from equation~\ref{eq:spatial_temporal_entropy_rate} can be meaningful. The differences between FSD and all other datasets could be attributed to the violation of one or both of these assumptions, due to the lower data quality. This impact underlines the importance of applying standard feature sets according to the assumptions they were derived under. Examining the other pairwise differences, we note that $C5$, corresponding to marginal dwell time, is most often able to differentiate datasets; differentiating the TAXI dataset from all others, which is sensible as taxis would have substantially different dwelling patterns than individuals. $C5$ also differentiates S9 from VIC. $C1$ differentiates TAXI from VIC and S9, and S10. The idea that TAXI are different than students or cyclists in terms of marginal velocity is not unreasonable, but that they are instead differentiated by squared velocity, which is proportional to energy is interesting, and worthy of further investigation. $C2$ differentiates S9 and S10. Again the interpretation that squared dwell time is different between similar datasets is intriguing but not clearly interpretable. $C3$ occasionally differentiates datasets, but only when both $C1$ and $C5$ already differentiate them.

Features were ordered according to f\_classif function in scikit-learn package in terms of their ability to distinguish between datasets using an ANOVA-based metric. Based on these results, an SVM was built to demonstrate the utility of combinations of features and determine the extent to which each feature provided additional discriminatory power between datasets. Other classifiers such as random forest are potential models and could be applied and compared with SVM in the future. Further analysis which explored the potential combinations of features and their performance with different classifiers is also a potentially informative avenue for future research. SVM analysis showed increasing accuracy with additional features, further reinforcing the idea that these features encode different spatial behaviors, and aid as a whole in discrimination. A maximum accuracy score of 0.85 on test set was achieved after all nine features were added. While the SVM analysis is valid for the datasets described here, different stratifications of data might be best described by a different feature ordering. Confusion matrix analysis demonstrated that the SVM was able to assign participants to FSD, VAN and TAXI with excellent accuracy, but that there was sufficient overlap between the more similar datasets S9, S10 and VIC. While VIC had a reasonable true positive rate, it was confounded by false positives from S9 and S10. A more sophisticated machine learning algorithm may have been able to differentiate individuals better. The key finding from this analysis was that while the features are successful at differentiating populations, differentiating individuals within populations is likely only feasible for highly distinct mobility behaviours. The development of additional features could improve classification at the individual level.

The analysis described here has important implications for understanding spatial systems and behavior. By leveraging a standardized feature set several advantages can be accrued. First, by packing these features into easily used Python code, rapid high-level analysis of new datasets, or new stratifications of existing data are possible. Absent strong hypothesis about spatial behaviors inherent in the data, this quick analysis could provide initial insight to further data exploration. Second, these baseline features provide a common language for summarizing newly published spatial datasets to concisely summarize general trends within the data. Finally, these features provide a standard discrimination baseline against which other novel features or feature sets can be compared for discriminatory ability against different data. Taken together these implications describe an important ongoing contribution to the spatial analysis community.

The features described here should be able to be employed across a wide variety of spatial data, no matter whether the data is from human, animal, or virtual movement, subject to the assumptions about data quality inherent in the features, because the features describe properties of trajectories with little to no appeal to the context under which those trajectories were formed. With sufficient, and sufficiently dense data, these features can be used to describe stratifications of any location-derived data. Because of the strong theoretical foundation of each of these features, differentiation within a feature has phenomenological implications. However, the interpretation of what these differences are implying about novel data is up to the researcher, and not directly implied by the features themselves.

While this work constitutes an important contribution to the spatial analysis literature it does suffer from shortcomings that could be addressed in future work. The first shortcoming is the relatively shallow analysis of the phenomenological implications of each feature. While each feature has been previously described in the literature in some form, exhaustive and comprehensive characterizations of the response of these features to different demographic and city specific mobility behaviours has not been conducted. This is a rich vein of research that should be explored. Each feature could also further benefit from characterization against further data, data collection techniques, and the impact of parameter choice. For example, the growth of buffer area with increasing buffer size might increase in proportion to the fractal dimension as more fractal paths experience increased buffer overlap with increasing buffer size. Due to limitations of GPS data which does not directly include knowledge of the environment, we can only infer the "when" and "where" of human movement, but cannot directly model why people create those trajectories. This requires joint analysis of GPS data with Geographic Information Systems, direct observations, surveys, or qualitative interviews. We have constrained our analysis to a relatively limited set of human movement patterns to demonstrate the concept. Understanding how these features interact with diverse data from human or possibly animal movement would help scope the utility of the approach. Although the SVM model shows the discriminatory power of feature combinations, additional research into their utility together using different feature ranking metrics and classifiers would be meaningful. Finally, amd perhaps most importantly the features described here are only a start and have known shortcomings. Many other spatial features such as concave hull are possible, and a comparison of features' different parameters such as the buffer distance of buffer area are worth exploring. The addition of novel features which capture other trajectory and spatial properties is desirable. A growing library of validated, phemenologically meaningful features would aid the community immensely.

\section{Conclusion}
In this paper we have described a standardized feature set for spatial analysis of location traces which employs a total of nine features drawn from three distinct mathematical disciplines: geometry, information theory, and fractal analysis. These features were evaluated for their ability to discriminate between six different GPS datasets of human mobility captured in four cities, each with a different core demographic. While no single feature was able to distinguish between each of the datasets, individually each was able to discriminate between a different subset of the datasets, and combined in an SVM was able to distinguish datasets with an accuracy of 0.85. This standard feature set will enable rapid exploration and standard analysis and reporting of data, making it faster, easier and more reliable to understand data collected for modeling and understanding spatial behavior.




%
\begin{acks}
We would like to ackowledge William van der Kamp for help with data access, Tuhin Paul for software support and NSERC and CIHR for funding.
\end{acks}

%
\bibliographystyle{ACM-Reference-Format}
\bibliography{sample-base}

%

\end{document}